# Ultra-high terahertz index in deep subwavelength coupled bi-layer free-standing flexible metamaterials

Leena Singh,[1] Ranjan Singh,[2,a)] and Weili Zhang [1,b)]

[1]School of Electrical and Computer Engineering, Oklahoma State University, Stillwater, Oklahoma 74078, USA

[2]Centre for Disruptive Photonic Technologies, Division of Physics and Applied Physics, School of Physical and Mathematical Sciences, Nanyang Technological University, Singapore 637371, Singapore

**Abstract**

We report extensive enhancement in the refractive index of artificially designed metamaterials by exploiting the deep subwavelength coupling in a free-standing, thin-film metal-dielectric-metal checkboard structure. A record high refractive index of 77.02+43.22$i$ is obtained at terahertz frequencies. The detailed investigations reveal that the enhancement of the effective refractive index of the structure via deep subwavelength coupling in a bilayer design is governed by a power law, which is an effective and simpler approach to design high index metamaterial. The approach relies on deep subwavelength coupling to obtain extremely high refractive index values that can lead to many practical applications in the field of imaging, lithography, design of delay lines and interferometers.

[a)]ranjans@ntu.edu.sg
[b)]weili.zhang@okstate.edu

## I. INTRODUCTION

IN a sense, every material is a composite, even if the individual ingredients consist of atoms and molecules.[1] These natural atoms when replaced with artificially designed unit cell structures (meta-atoms), result in an engineered material that is now well known as metamaterials. Metamaterial properties are derived from their meta-atoms rather than their composition. These sub-wavelength meta-atoms can be designed with unique shapes and sizes, and wit-fully arranged and oriented to achieve unconventional values of permeability[2] and permittivity[3], the characteristic electromagnetic properties of any material medium. By tailoring the electric and magnetic response of the material towards incident electromagnetic waves, the effective refractive index can be varied from negative[4-9] to zero,[10-13] or even to higher positive values.[14-20] Metamaterials have created a completely new dimension to engineered materials. This has offered an opportunity to build new devices with exotic functionalities.[21] Metamaterials are proven to be especially valuable in the terahertz regime, where most naturally existing materials exhibit weak electromagnetic wave response.[22-24]

Recently, high-index metamaterials have drawn much attention.[17] Higher values of positive refractive indices are especially desired in the field of imaging and lithography, where the resolution scales inversely with the refractive index.[25] Increasing the refractive index over a large frequency range results in the broadband slow light, which can be used to enhance the storage capacity of the delay lines.[26] High indexed metamaterials may also enhance the spectral sensitivity of certain types of interferometers[27] and can benefit many other practical application areas. Most of the previous high indexed metamaterials offer a specific value of fixed high refractive index or short range of high index values.[20] Enhancing refractive index values for any of these previous structures usually involves dealing with complicated parameter trade-off equations or structural element changes.[14,15] Another method used so far, to increase refractive index values is to stack multiple layers of same structure, which is not only a complicated approach due to the alignment issues but also results in thicker metamaterials, and hence limits the maximum values of refractive index that can be achieved. Enhancing the effective value of refractive indices over a broad range, for a thin film, free standing metamaterial, still remains a challenge. A performance (effective refractive index values) comparison of the proposed structure with that of the current state of the art is presented in Table 1.

In this article, we propose and demonstrate a high index metamaterial that consists of aluminum closed resonator array on two sides (front and back) of a mechanically stable, thin-film, free-standing polyimide substrate. Figure 1a shows the microscopic image of the checkboard structure. The structure is symmetrical in both the $x$- and $y$-directions, making the metamaterial insensitive to the polarization of incoming light at normal incidence. The inset of Figure 1a shows the design and dimensions of the unit cell structure of the metamaterial. The length of the metallic unit cell structure is set at 60 μm and the width of the metal bars, $w$ is 5 μm. The entire sample consists of 156 × 156 unit cells with a periodicity of 64 μm.

## II. EXPERIMENTAL AND NUMERICAL DESIGN

To investigate the effect of deep subwavelength coupling in the proposed metamaterial structure, two samples with polyimide substrate thickness of 22 μm and 1 μm were fabricated and characterized using terahertz time-domain spectroscopy (THz-TDS). Two layers of polyimide (PI 2525) 11 μm each, were spin coated on a silicon substrate by conventional lithography techniques, each followed by soft bake for 2 minutes at 120º C. The final 22 μm thick polyimide spacer layer was cured at 150º C, with an increase of 5º C per minute up to 250º C, which was maintained to be constant for 30 minutes and then decreased to 100º C for 20 minutes. A single layer of polyimide (PI 2610) was spin coated at 500 and 5000 rpm for 5 and 30 seconds, respectively, followed by a soft bake at 90º C for 90 seconds. The 1 μm thick polyimide spacer layer was then heat cured with an increase of 4º C per minute from room temperature, up to 350º C, which was maintained for 30 minutes. Both the polyimide layers were finally allowed to be cooled down to room temperature. An aluminum layer of 0.2 μm was uniformly deposited on the cured polyimide layers, by thermal evaporation, which was patterned using positive photoresist, PR1-4000A. The polyimide layer with metal structure was then peeled off from the silicon wafer to form a free standing sample. Another 0.2 μm thick aluminum layer was deposited and structured on blank side of the polyimide substrate. Finally, the fabricated thin-film, free-standing metamaterial samples are shown in Figure 1b.

A photoconductive switch based THz-TDS system that consists of four parabolic mirrors, arranged in an 8*f* confocal geometry was used to characterize the transmission properties of the metasurface.[28-30] The transmission is recorded with the terahertz beam illuminating the sample at normal incidence, along the *z*-axis. The electric field vector is in the *x*-direction and the magnetic field vector is in the *y*-direction (*x, y* and *z* in Figure 1a). The *8f* confocal system illuminates the metasurface sample with a focused terahertz beam of 3.5 mm waist diameter. The transmission amplitude is calculated as

$$t(\omega) = \left| Es(\omega) \Big/ Er(\omega) \right|, \qquad (1)$$

where $Es(\omega)$ and $Er(\omega)$ are the Fourier transform of the transmitted signal through the sample and the reference, respectively. The free space terahertz pulse in dry nitrogen purged environment is used at normal room temperature as a reference. The experiment is conducted with 17 ps time-domain terahertz pulse scan, so as to provide a spectral resolution of 59 GHz to the measurements.

The measured transmission amplitude through the metamaterial sample *(orange)* and air as reference *(pink)* are shown in Figure 2a, for both 22 μm and 1 μm samples, respectively. The transmission through the 22 μm sample is quite significantly attenuated beyond 1.12 THz and remains almost zero in the frequency band of 1.2-2.1 THz. An extensive set of simulations have been performed using commercially available Finite Integration Method solver in CST microwave studio.

During simulations, the sample shows an angle independent behavior, hence the angle of incidence was kept normal to the sample surface. The background material properties were set to type normal, i.e. air with $\varepsilon = 1.0$ and $\mu = 1.0$. Closed electric and magnetic boundary conditions were set along the *x*-axis and *y*-axis, respectively, while open boundary condition was employed along the *z*-axis. The observation ports were set at a distance of 5 μm along the *z*-axis from the sample surface, in order to obtain far field response for the *x*-polarized incident wave. Polyimide (lossy) with permittivity ($Re[\varepsilon] = 2.96$)[15] and aluminum with conductivity $2.2 \times 10^7$ S/m[31] available with the CST material library are used in the simulation. Experimentally measured *(pink)* and the simulated *(blue)*, transmission amplitudes are found to be in excellent agreement with each other, as shown in Figure 2b, for both 22 μm and 1 μm samples, respectively. The slight deviations arise from the non-conformities during the fabrication process and variations in the material parameters used for the simulations.

## III. RESULTS AND DISCUSSION

The simulated reflection parameter, S11 (green) and the transmission parameter, S21 (blue) for the 22 μm and 1 μm samples are shown in Figure 3a. For the 22 μm sample, the transmission is drastically suppressed after 1.12 THz, and the reflection reaches unity and in the case of the 1 µm sample the reflection gradually approaches to unity near 1.5 THz. This clearly shows that most of the incident wave energy is reflected back rather than being scattered at non-zero angles with respect to the surface normal. The checkboard structure can be seen as an infinite periodic grating, in reference to diffraction theory. Since the unit cell of periodicity 64 µm is much smaller as compared to the wavelength of interest (~299 µm), hence the first order diffraction for the structure does not exist at normal incidence.

Standard parameter retrieval equation was adopted for refractive index extraction using the simulated transmission (S21) and reflection (S11) values.[32] In order to choose the correct branch of the real part of the refractive index,[33] a customized matlab code[34] is used to extract the complex refractive index of the metamaterial solving the following equations,[33]

$$S_{11} = \frac{R_{01}\left(1-e^{i2nk_0 d_{eff}}\right)}{1-R_{01}^2 e^{i2nk_0 d_{eff}}}, \qquad (2a)$$

$$S_{21} = \frac{(1-R_{01}^2)e^{ink_0 d_{eff}}}{1-R_{01}^2 e^{i2nk_0 d_{eff}}}, \qquad (2b)$$

where $R_{01} = \frac{Z-1}{Z+1}$,

$d_{eff}$ is the effective thickness of the homogeneous slab. In order to consistently estimate the effective refractive index, the sum of substrate thickness, $t$ and twice of metal layer thickness, $t_m$ (for two layers) is used as $d_{eff}$, effective thickness, for equation 2a and 2b. Since the experimental and the simulated transmission amplitudes are in excellent agreement with each other, the simulated transmission (S21) and reflection (S11) values can be reliably used for the estimation of the effective parameters.

The calculated effective refractive index of the medium, with the highest peak value of 7.78+0.96$i$ at 0.839 THz for the 22 µm sample and 41.8+22.07$i$ at 2.026 THz for the 1 µm sample is shown in Figure 3b. When there is a decrease of 95.5% in the thickness of the sample, the real part of the refractive index shows an increase of 437.3%. It is interesting to note that although the imaginary part of the refractive index, which is indicative of the absorption losses in the sample, shows an increase of 2203.76%, with the same decrease in sample thickness, the value of the imaginary part is smaller as compared to their real counterparts, hence it does not significantly hamper the performance of the samples. The smaller value of imaginary part of the refractive index as compared with their real counterparts, for both the 22 μm and 1 μm samples, at the respective frequencies of interest is indicative of lower absorption losses in the metamaterial.

It is worth noting that the effective homogeneity condition, which requires that the thickness along the propagation direction is smaller than the operating wavelength, remains valid throughout the frequency band of operation. [15]

The effective permittivity and permeability of metamaterial samples are numerically extracted using the following equations, [33]

$$\varepsilon_{eff} = \frac{N_{eff}}{Z_{eff}}, \quad (3a)$$

and

$$\mu_{eff} = N_{eff} \cdot Z_{eff}, \quad (3b)$$

where $N_{eff}$ and $Z_{eff}$ are the effective refractive index and normalized impedance retrieved from the S-parameters, respectively.[15]

The real part of the effective refractive index *(red)* for 22μm samples shown in Figure 3b, shows two peaks. The first peak value of the real part of the effective refractive index is due to permeability peak of 9.75+7.69$i$ at 0.827 THz and the second local maxima of the real part of the effective refractive index, is contributed by the effective permittivity peak of 287.6+198.8$i$, at 1.623 THz, as shown in Figure 4a. This clearly shows that the peak effective refractive index of the 22 μm sample is contributed by the high magnetic activity [35-37] of the structure at resonance frequency. Similarly, for the 1 µm sample, the first local maxima of the real part of the effective refractive index (red), as shown in Figure 3b, is due to the permeability peak of 2.51+1.484$i$ at 0.967 THz, as shown in Figure 4b, while the second maxima, which also happens to be the peak value of the real part of the effective refractive index, is contributed mainly by the permittivity peak of 2336+1993$i$ at 1.99 THz and not by the permeability, which remains unity at the frequency of interest, as shown in Figure 4b.

For both 22 µm and 1 µm samples, we observe two peaks in the real part of the effective refractive index values. The first peak appears due to high permeability of the sample and the second peak is contributed mainly by the high permittivity of the sample. In the case of the 22 µm sample, the highest value of the real part of the effective refractive index (first peak) appears due to high permeability of the sample, while in the case of the 1 µm sample, the highest values of the real part of the effective refractive index (second peak) appears due to the high permittivity of the sample. This exceptionally high value of the effective permittivity, in the case of the 1 µm sample is in fact, the result of the deep subwavelength coupling between the meta-atoms on the opposite sides of the structure, which is enhanced due to the decrease in the substrate thickness.

Moreover, it is noteworthy that in Figure 4a and b, the imaginary parts of both permittivity and permeability attain negative values for different frequencies. This is not in contradiction with any physical law. [38-42] The dissipated energy, $W$ is given by the following equation, [43]

$$W = \frac{1}{4\pi} \int d\omega \omega [\varepsilon''|E(\omega)|^2 + \mu''|H(\omega)|^2], \qquad (4)$$

where $\varepsilon''$ and $\mu''$ are the imaginary parts of the effective permittivity and permeability, respectively. The condition W>0, does not require that $\varepsilon''$ and $\mu''$ must be simultaneously positive.[44] The antiresonant behavior is caused by the requirement that the refractive index must be bounded in the structures which possess finite spatial periodicity.[43]

We further investigate the distribution of electromagnetic field inside the structure, the electric and magnetic energy density at 0.839 THz for the 22 μm sample and at 2.026 THz for the 1 μm sample is shown in Figure 5. For the *x*-polarized incident electromagnetic wave, the electric energy is mainly concentrated on the edges of the structure along the *y*-axis, which shows a strong coupling between the adjacent structures on the same side of the substrate. The magnetic energy density inside the structure is very weak, other than central metal bar along the *x*-axis. Owing to the thin metal layer of 200 nm and 5 μm width, the area subtended by the induced current loop is very small, which does not support a high magnetic moment and hence, leading to an extremely low diamagnetic effect at the frequency band of interest.[15-17] The permittivity of metamaterials, which is mainly due to the strong capacitive coupling between the adjacent unit cell, can be increased multiple-fold, by decreasing the substrate thickness and hence allowing deep subwavelength coupling between the meta-atoms on the opposite sides of the structure.

In order to explore the enhancement of the effective refractive index by means of deep sub-wavelength coupling, an extensive study of six samples with varying dielectric spacer thickness has been carried out. All the geometric parameters except for the dielectric thickness (*t*) are kept constant for all the samples. The effect of the change in thickness, *t* of the polyimide spacer layer on the effective refractive index of the metamaterial is shown in Figure 6a. Refractive index as high as 77.02 at 2.11 THz is achieved by reducing the polyimide spacer thickness to 100 nm. This value of refractive index is five-fold higher than that of the initial structure, with 22 μm thick spacer layer. The numerical simulations demonstrate, that as the spacer thickness decreases, the deep subwavelength coupling between the meta-atoms on the opposite side of the spacer layer is enhanced which leads to an increase in the effective refractive index of the sample.

The numerically calculated peak value of the effective refractive indices as a function of the dielectric spacer thickness is shown in Figure 6b. To our surprise, the peak value of the effective refractive indices ($N_{peak}$) demonstrates a universal power law behavior, with respect to the dielectric spacer thickness ($t$) described by

$$N_{peak} = 37.076 t^{-0.424}, \qquad (5)$$

This is further verified by plotting the natural log of both spacer thickness and the peak value of the effective refractive indices shown in Figure 6c. The data could be easily fit to a straight line, which is a signature of the power law and is indicative of the scale invariance and the universality of the function,

$$\ln(N_{peak}) = -0.4243 \ln(t) + 3.613, \qquad (6)$$

The above equation clearly establishes, that exploiting the deep subwavelength coupling between the meta-atoms on the opposite side of the substrate by tailoring the dielectric spacer thickness, is one of the most straightforward and efficient ways to enhance the peak refractive index over a much wider range of values. The significance of this finding lies in the ease and capability to use a rather simple structure, to achieve the precise values of the positive high refractive index, as required in various fields of application.

The effect of deep sub-wavelength coupling can be clearly visualized by the surface current distribution in the structure. Figure 7 shows the surface current distribution for the 22 μm and 1 μm thick samples. The metal structure on the back side of the 22 μm thick substrate (that acts as the spacer) is weakly coupled with that of the front side metal closed ring, as seen by the incident wave, leading to lower permittivity and hence lower effective refractive index peak as compared to that of thinner substrates. For the 1 μm spacer thickness, there exists a significantly strong subwavelength coupling between the opposites sides of the substrate, as illustrated in Figure 7 by the stronger surface current distribution.

Another set of simulations was carried out, in order to study the effect of alignment between the two sides of the structure, on the peak refractive index value. It was observed that the peak refractive index value which depends on the strength of the subwavelength coupling between meta-atoms at the two sides of the sample, as visualized by the surface current distribution in Figure 7, in turn depends on the amount of the area of the metallic elements of the two sides being superposed on each other. This finding opens up another opportunity for further enhancement of the refractive index, by changing the structure, so as to wit fully increase this superposition area between the two sides.

We further investigated the effect of the change in the refractive index of the spacer layer to enhance the deep subwavelength coupling between the two metal sides of the structure. In order to systematically examine the effect of the change in the substrate refractive index on the effective refractive index ($N_{eff}$) of the overall structure, two samples with 22 μm and 1 μm thicknesses were simulated and studied with varying substrate refractive index, keeping all other dimensions of the structure constant. The effective refractive index of the structures increases with increasing the substrate refractive index, as shown in Figure 8a and 8b. The higher refractive index substrates further facilitated the coupling between the meta-atoms on opposite sides of the samples. The plot of the peak value of the effective refractive index ($N_{peak}$) as a function of the spacer layer refractive index ($N_{spacer}$) for 22 μm and 1 μm samples, shown in Figure 8a and 8b, perfectly follows linear behavior, and is expressed as Equations 7 and 8, respectively,

$$N_{peak} = 3.5052 N_{spacer} + 1.6309, \qquad (7)$$

$$N_{peak} = 6.5148 N_{spacer} + 29.951. \qquad (8)$$

This finding is significant, as it provides the capability to further tailor the effective refractive index of the metamaterial, independent of the thickness of the substrate. Moreover, the fact, that Equation 8 for the 1 μm sample, shows a steeper slope as compared to that of Equation 7 for the 22 μm sample, indicates that increasing the spacer refractive index is more effective when the sample thickness is low. Theoretical and numerical simulations reveal that by combining the effect of the both, decreased substrate thickness, to 0.1 μm and increased substrate refractive index, of 5 units, a peak refractive index of 83.42 at 1.93 THz, can be achieved.

**IV. CONCLUSION**

We have experimentally and numerically demonstrated the effect of enhancing the deep subwavelength coupling to achieve extremely high refractive index in terahertz metamaterials. More than five-fold enhancement in the peak value of the effective refractive index was realized by merely altering the dielectric spacer thickness. As a result, a peak refractive index of about 77.02 at 2.11 THz was experimentally characterized. Exploiting deep subwavelength coupling, by means of varying dielectric spacer thickness would benefit other metamaterials structures as well, to enhance the peak values of their refractive index. Such designs of high refractive index metamaterials may find applications in terahertz photonics where the naturally occurring materials have limited response.

**ACKNOWLEDGMENTS**

This work was partially supported by the National Science Foundation (Grand No. ECCS-1232081).

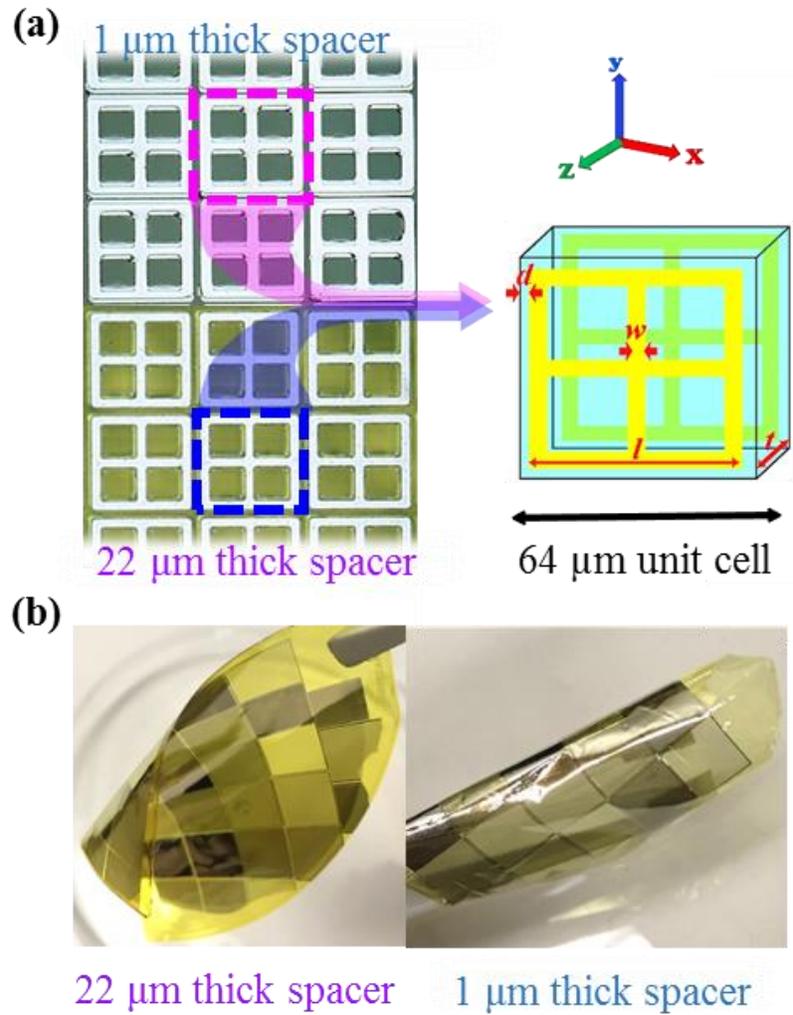

**Figure 1.** Microscopic images and Unit cell design. (a) Microscopic image of the fabricated samples(left). Dimensions of the checkboard structure are $l$ = 60 μm, $w$ = 5 μm, $d$ = 4 μm and $t$ = 22 μm and 1 μm, respectively. The aluminum structures of thickness, $t_m$ = 0.2 μm are fabricated on both the sides of the polyimide substrate (spacer layer), 3D view shown on the right side. (b) Photograph showing flexibility of the fabricated free standing samples, with 22 μm and 1 μm substrate thickness, respectively.

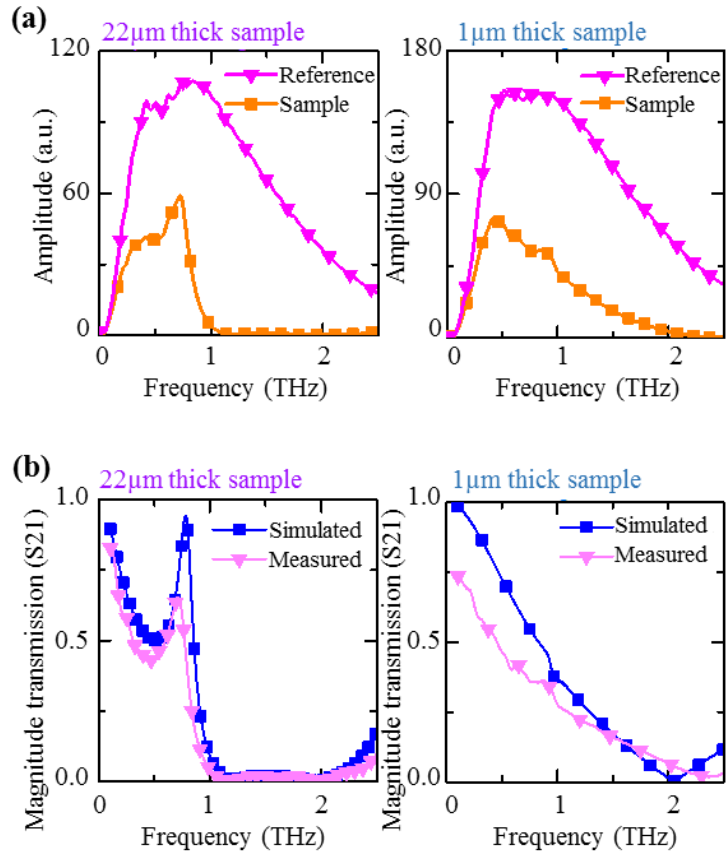

**Figure 2.** Simulated and measured transmission map. (a) Measured transmission amplitude for the 22 μm and 1 μm thick samples is compared with that of the reference, air and (b) comparison between the simulated and the measured transmission magnitude for the samples is shown for the 22μm and 1 μm thick samples, respectively.

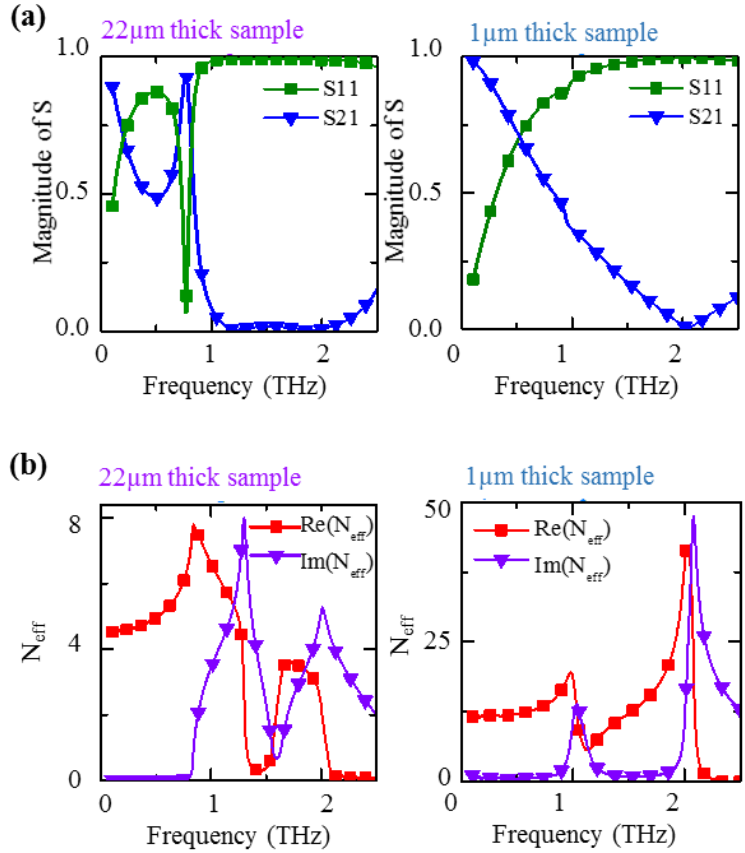

**Figure 3.** Simulated S-parameter and calculated refractive index. (a) Simulated reflection, S11 (green) and transmission, S21 (blue). (b) Calculated complex refractive index of the samples with peak values 7.78+0.96$i$ at 0.839THz and 41.8+22.07$i$ at 2.026 THz for the 22 μm and 1 μm samples, respectively.

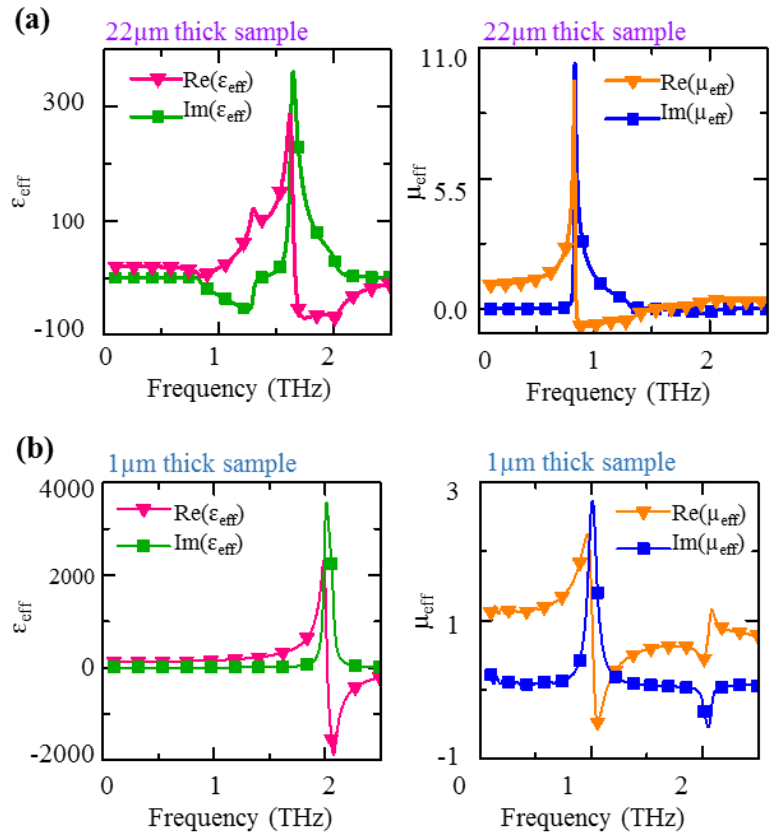

**Figure 4.** Calculated permeability and permittivity. (a) and (b) Numerically obtained values of effective permeability and permittivity from the S-parameter retrieval method for the 22 μm and 1 μm samples, respectively.

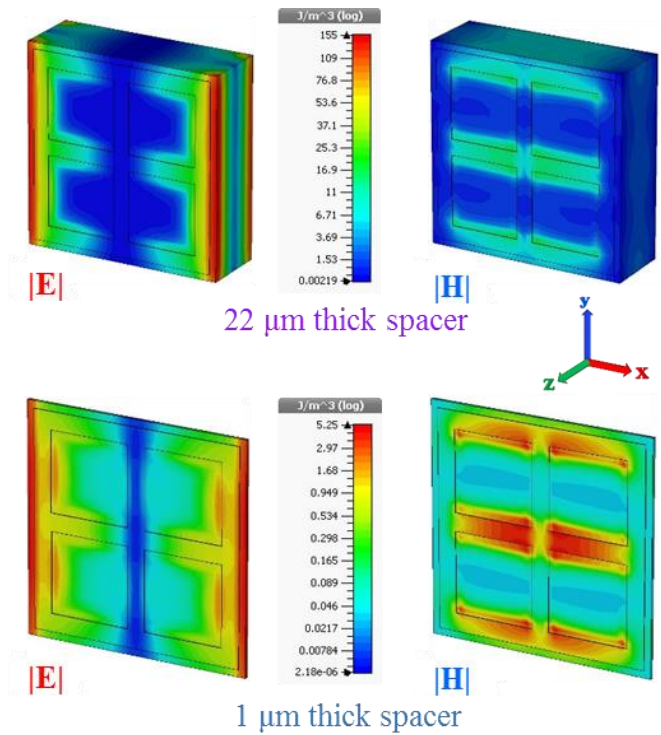

**Figure 5.** Energy density. Simulated distribution of the electric and magnetic energy density at 0.839 THz and 2.026 THz for the 22 μm and 1 μm samples, respectively.

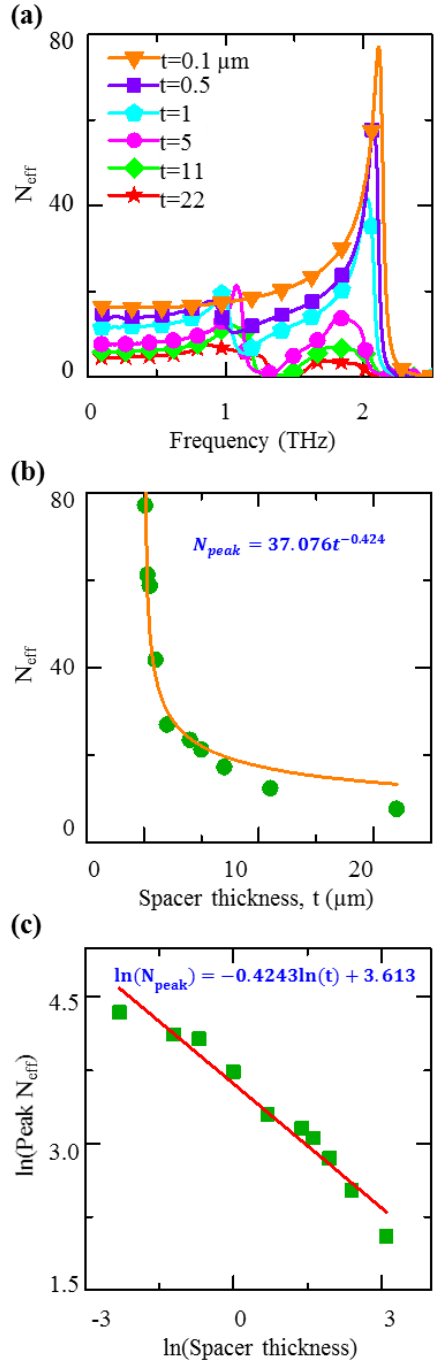

**Figure 6.** Effective refractive index versus substrate thickness. (a) Calculated effective refractive index for the various spacer thicknesses, (b) change in the peak value of the effective refractive index w.r.t. the spacer thickness and (c) natural log of the peak refractive index versus natural log of the spacer thickness.

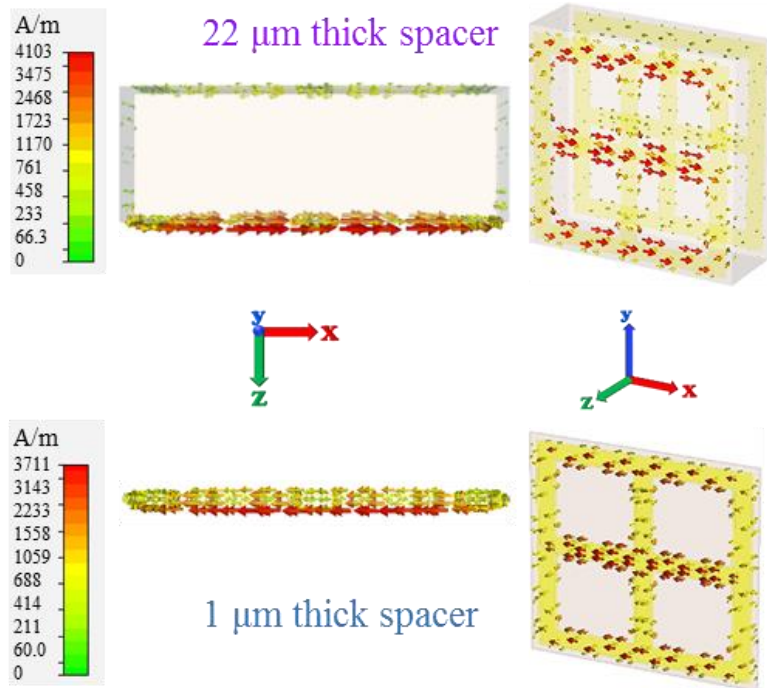

**Figure 7.** Surface current distribution. Top and perspective view of the surface current distribution for normal incident wave along the *z*-direction inside the structure with (a) 22 μm thick substrate and (b) 1 μm thick substrate.

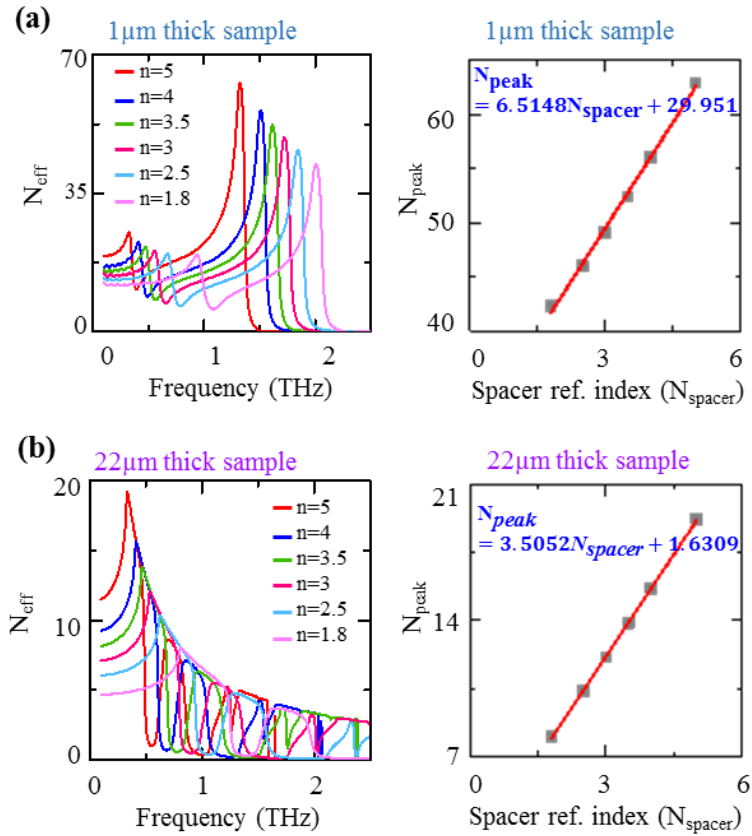

**Figure 8.** Effective refractive index versus substrate refractive index. Calculated effective refractive index for varying the spacer refractive index for (a) 1 μm and (b) 22 μm thick sample. Change in the peak value of the effective refractive index of the sample w.r.t. the spacer refractive index for (a) 1 μm and (b) 22 μm thick sample.

**Table 1.** Performance comparison. The effective refractive index value of the current state of the art are compared with that of the proposed structure.

| Reference article | Frequency | Highest value of ref. index |
|:---:|:---:|:---:|
| Ref. 13. | 0.384 THz | 61.83 |
| Ref. 14. | 0.333 THz, | 14.36 |
| Ref. 16. | 0.315 THz | 54.87 |
| This work | 2.11 THz | 77.02 |